\documentclass[conference]{IEEEtran}
\IEEEoverridecommandlockouts
\usepackage{cite}
\usepackage{amsmath,amssymb,amsfonts}
\usepackage{algorithmic}
\usepackage{graphicx}
\usepackage{textcomp}
\usepackage{xcolor}
\usepackage{enumerate}
\usepackage{booktabs}
\usepackage{multirow}

\def\BibTeX{{\rm B\kern-.05em{\sc i\kern-.025em b}\kern-.08em
    T\kern-.1667em\lower.7ex\hbox{E}\kern-.125emX}}
\begin{document}


\title{The Virtuous Cycle: AI-Powered Vector Search and Vector Search-Augmented AI}

\author{
\IEEEauthorblockN{Jiuqi Wei}
\IEEEauthorblockA{
\textit{Oceanbase, Ant Group}\\
Beijing, China\\
weijiuqi.wjq@antgroup.com}
\and
\IEEEauthorblockN{Quanqing Xu}
\IEEEauthorblockA{
\textit{Oceanbase, Ant Group}\\
Hangzhou, China\\
xuquanqing.xqq@oceanbase.com}
\and
\IEEEauthorblockN{Chuanhui Yang$^\ast$}
\IEEEauthorblockA{
\textit{Oceanbase, Ant Group}\\
Beijing, China\\
rizhao.ych@oceanbase.com}
\thanks{*Chuanhui Yang is the corresponding author.} 
}

\maketitle

\begin{abstract}
Modern AI and vector search are rapidly converging, forming a promising research frontier in intelligent information systems.
On one hand, advances in AI have substantially improved the semantic accuracy and efficiency of vector search, including learned indexing structures, adaptive pruning strategies, and automated parameter tuning. 
On the other hand, powerful vector search techniques have enabled new AI paradigms, notably Retrieval-Augmented Generation (RAG), which effectively mitigates challenges in Large Language Models (LLMs) like knowledge staleness and hallucinations.
This mutual reinforcement establishes a ``virtuous cycle" where AI injects intelligence and adaptive optimization into vector search, while vector search, in turn, expands AI's capabilities in knowledge integration and context-aware generation.
This tutorial provides a comprehensive overview of recent research and advancements at this intersection.
We begin by discussing the foundational background and motivations for integrating vector search and AI.
Subsequently, we explore how AI empowers vector search (AI4VS) across each step of the vector search pipeline.
We then investigate how vector search empowers AI (VS4AI), with a particular focus on RAG frameworks that integrate dynamic, external knowledge sources into the generative process of LLMs.
Furthermore, we analyze end-to-end co-optimization strategies that fully unlock the potential of the ``virtuous cycle" between vector search and AI.
Finally, we highlight key challenges and future research opportunities in this emerging area.
\end{abstract}


\section{Introduction}

The past decade has witnessed two revolutions in information science: vector search, which empowers machines with human-like semantic comprehension for information retrieval and recommendation, and modern artificial intelligence (AI), particularly the rise of large language models (LLMs), which provides machines with unprecedented capabilities in generation and reasoning.
Collectively, these advances are fundamentally reshaping the way humans interact with vast information resources.

Nowadays, vector search and AI are no longer parallel fields; they are increasingly converging and forming a cutting-edge research frontier.
On one hand, advances in AI empower vector search with intelligence and adaptive optimization, leading to substantial enhancements in search accuracy and efficiency. 
Notable progress has been made through learned indexing structures, adaptive pruning strategies, and automated parameter tuning.
On the other hand, powerful vector search enhances AI’s capacity for knowledge integration and context-aware generation. 
This has given rise to new paradigms such as Retrieval-Augmented Generation (RAG), which effectively mitigates challenges in Large Language Models (LLMs) including knowledge staleness and hallucination.
This mutual reinforcement establishes a ``virtuous cycle" where AI advances vector search (AI4VS) and vector search, in turn, augments AI (VS4AI).
This tutorial systematically analyzes this bidirectional relationship, delineating how AI optimizes vector search processes and vice versa.

\begin{figure}[tb] 
	\centering
	\includegraphics[width=\linewidth]{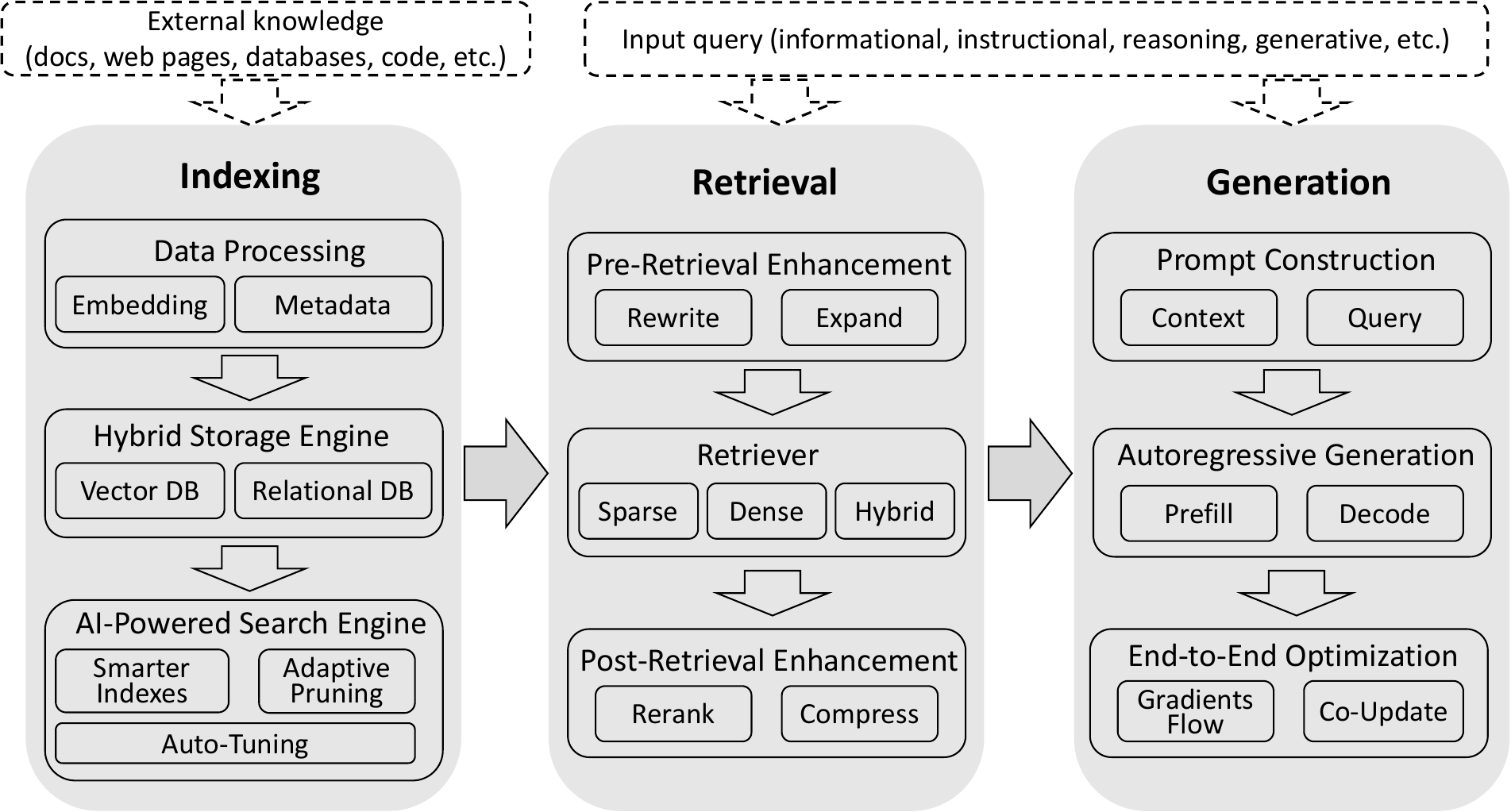}
	\caption{An architecture of AI-Powered Vector Search (AI4VS), Vector Search-Augmented AI (VS4AI), and their End-to-End Optimization.}
	\label{fig:framework}
\end{figure}

\textbf{Tutorial Overview.} 
This tutorial provides a comprehensive overview
of the latest research and advancements at the intersection of vector search and AI.
Figure~\ref{fig:framework} provides an architecture of this intersection.
The tutorial is designed to last \textbf{1.5 hours} and is structured into five parts:
\begin{enumerate}
  \item \textbf{Background and Motivation (25 mins).}
  This part introduces the fundamentals of vector search and generative AI, and explain why vector search needs AI and vice versa by analyzing their respective challenges.
  \item \textbf{AI-Powered Vector Search (25 mins).}
  This part explores how AI empowers vector search (AI4VS) across each step of the vector search pipeline, including learned indexing structures, adaptive pruning strategies, and automated parameter tuning.
  \item \textbf{Vector Search-Augmented AI (25 mins).}
  This part investigates how vector search empowers AI (VS4AI), with a particular focus on the evolution of the RAG framework: from Naive RAG to Advanced RAG and finally to Modular RAG.
  \item \textbf{Closing the Loop: End-to-End Optimization (10 mins).}
  This part analyzes end-to-end co-optimization strategies that fully unlock the potential of the ``virtuous cycle" between vector search and AI.
  \item \textbf{Challenges and Opportunities (5 mins).}
  This part highlights key challenges and future research directions in this emerging area.
\end{enumerate}

\textbf{Target Audience.} 
This tutorial is intended for students, researchers, and developers who are interested in vector search, AI, and their intersection.
The tutorial provides a comprehensive and self-contained introduction that does not require specific prerequisite knowledge.

\section{Tutorial Outline}

\subsection{Background and Motivation}

We begin by introducing the fundamentals of vector search and generative AI, and explain why vector search needs AI and vice versa by analyzing their respective challenges.

\subsubsection{\textbf{Overview of Vector Search}} Vector search can be systematically introduced through three core aspects: foundational problems, technical methods, and real-world systems.

\indent \underline{\emph{Problems.}}
We first present the fundamental problems underlying vector search: Approximate Nearest Neighbor Search (ANNS), Maximum Cosine Similarity
Search (MCSS), and Maximum Inner Product Search (MIPS), with the mathematical relationships and conversion methods between them.

\underline{\emph{Methods.}}
Next, we propose a taxonomy of state-of-the-art vector search methods, categorizing them into locality-sensitive hashing methods~\cite{dblsh,pmlsh,detlsh},
subspace collision methods~\cite{wei2025subspace},
vector quantization methods~\cite{jegou2010product,ge2013optimized,gao2024rabitq}, tree methods~\cite{annoy,dasgupta2008random,ads}, graph methods~\cite{munoz2019hierarchical,fu2019fast,malkov2018efficient}, and hybrid methods~\cite{chen2018sptag,azizi2023elpis,gou2025symphonyqg}, and conduct a systematic analysis of their respective strengths and limitations.

\underline{\emph{Systems.}} 
Finally, we introduce the architectures of modern vector databases, such as FAISS~\cite{douze2025faiss}, Milvus~\cite{wang2021milvus}, ScaNN~\cite{guo2020accelerating}, and VSAG~\cite{zhong2025vsag}, which combine multiple vector search techniques for scalability and latency optimization.

\subsubsection{\textbf{Challenges of Vector Search}} 
\label{sec:challenge_vs}
Traditional vector search faces a range of challenges that can be effectively mitigated using modern AI techniques (solutions in Section~\ref{sec:ai4vs}).


\indent \underline{\emph{Inaccurate index.}}
Traditional index structures rely on expert experience and heuristic design rules, making them difficult to adapt to different data distributions and unable to effectively handle the division of boundary areas. 
Therefore, building smarter indexes for millions or even billions of data, so that the index structure can be dynamically adjusted according to the data distribution, becomes a key challenge.

\indent \underline{\emph{Inefficient Query.}}
Traditional vector search methods adopt fixed index traversal strategies for all queries and cannot dynamically adjust computing resources based on the characteristic (e.g., easy or hard) of the current query, resulting in a waste of query overhead. 
Therefore, designing adaptive pruning that intelligently determine the search path and depth during search to avoid unnecessary traversal becomes a key challenge.

\indent \underline{\emph{Non-optimal parameters.}}
A vector search system involves many parameters, requiring careful tuning from index construction to query answering. 
Traditional parameter tuning relies on expert-driven manual trial, which is costly, non-optimal, and incapable of adapting to dynamic data shifts. 
Therefore, achieving automated parameter tuning becomes a key challenge.

\subsubsection{\textbf{Overview of Generative AI}} 
A systematic introduction to Generative AI can be structured around three key techniques: the Transformer architecture as the foundational model, the self-attention mechanism that enables its contextual understanding, and LLMs as the scaled-up realization.

\underline{\emph{Transformer.}} 
We begin by introducing the Transformer architecture~\cite{vaswani2017attention} as the fundamental building block of modern generative AI.
We discuss key architectural variants, including encoder-only~\cite{devlin2019bert}, decoder-only~\cite{achiam2023gpt}, and encoder-decoder~\cite{raffel2020exploring}, and illustrate application tasks for each variant.


\underline{\emph{Self-attention.}} 
Next, we delve into the mathematical formulation and computational paradigm of the self-attention mechanism~\cite{vaswani2017attention}. 
We detail how the Query-Key-Value mechanism operates and its role as the source of Transformer's contextual understanding power.

\underline{\emph{LLMs.}} 
Finally, we introduce LLMs as the scaled-up realization of the Transformer architecture. 
The discussion further covers the evolution of LLMs and the state-of-the-art open-source~\cite{bai2023qwen,liu2024deepseek,touvron2023llama} and closed-source~\cite{achiam2023gpt,team2023gemini} models.

\subsubsection{\textbf{Challenges of Generative AI}} 
\label{sec:challenge_ai}
LLMs confront several challenges that can be effectively addressed through vector search techniques (solutions in Section~\ref{sec:vs4ai}).

\underline{\emph{Knowledge Staleness.}}
LLMs are trained on static corpora, lacking awareness of recent events or evolving information after training. 
As a result, their knowledge becomes outdated quickly and may not cover domain-specific or recent facts.

\underline{\emph{Hallucination.}} 
LLMs are prone to generating plausible but factually incorrect or completely fabricated content when facing knowledge gaps or ambiguous queries, because they rely on internal statistical patterns rather than verified evidence.

\underline{\emph{High Cost.}} 
Encoding more world knowledge into the model requires expanding parameters, which leads to a sharp increase in training and inference costs, and scaling model size does not guarantee better factual recall.

\underline{\emph{Lack of Specificity.}}
General-purpose LLMs are trained on the public corpora and struggle with specialized domains (e.g., legal, medical, enterprise-specific knowledge) or user-specific contexts without costly fine-tuning.

\subsection{AI-Powered Vector Search (AI4VS)}
\label{sec:ai4vs}
We explore how AI techniques leverage their powerful capabilities in intelligence and adaptive optimization to address the challenges of vector search outlined in Section~\ref{sec:challenge_vs}.



\subsubsection{\textbf{Smarter Indexes}}
\emph{Learning to hash} methods leverage deep neural networks to transform high-dimensional vectors into compact hash codes, preserving semantic similarities between the original and encoded spaces to enable efficient search in this compressed domain~\cite{li2020efficient,jiang2018asymmetric,lai2015simultaneous}.
\emph{Learning to partition} methods develop data-driven partitioning strategies that adapt to the underlying data distribution and cluster boundaries, thereby enhancing overall index accuracy~\cite{zeng2025lira,gupta2022bliss,li2023learning,dong2020learning}.
\emph{Vector quantization} methods use machine learning algorithms to compress high-dimensional vectors into compact quantization codes by minimizing reconstruction error or preserving neighborhood structures, enabling high-efficiency approximate search with controllable accuracy~\cite{jegou2010product,ge2013optimized,kalantidis2014locally}.

\subsubsection{\textbf{Adaptive Pruning}}
\emph{Early termination} strategies employ models to estimate the minimum number of vectors or clusters that must be visited to retrieve the ground-truth k-nearest neighbors or declarative recall, enabling query-aware search and reducing computational cost~\cite{li2020improving,chatzakis2025darth}.
\emph{Partition pruning} leverage models to predict the query-specific number of partitions required for effective search, dynamically restricting traversal to relevant subsets and minimizing unnecessary partition accesses~\cite{zheng2023learned,zeng2025lira}.
\emph{Route pruning} reframes the graph routing problem within a probabilistic modeling framework, learning an efficient routing policy to guide the search for nearest neighbors on proximity graphs~\cite{baranchuk2019learning}.
\emph{Refinement pruning} utilizes a learned orthogonal transform and progressively refines distance bounds during the refinement phase, enabling real-time candidate pruning whenever a candidate's lower bound surpasses the current best-distance threshold~\cite{ramani2025panorama}.

\subsubsection{\textbf{Automated Parameter Tuning}}
\emph{Search-based} methods operate by defining a vast configuration space and empirically evaluating numerous candidate parameter sets to identify the optimal configuration~\cite{zhou4925468auto,wang2021comprehensive}.
While thorough, this approach can be computationally intensive.
\emph{Learn-based} methods leverage machine learning models (e.g., Bayesian optimization, classifiers, or pre-trained meta-models) to intelligently recommend promising configurations that meet target recall thresholds or to directly predict the query performance of various parameter settings~\cite{yang2024vdtuner,duan2025pgtuner,oyamada2020towards}.
This paradigm significantly reduces the need for exhaustive manual testing.
\emph{Hybrid} methods integrate the robustness of search-based methods with the efficiency of learn-based approaches, paving the way for more adaptive and self-optimizing vector databases~\cite{zhong2025vsag}.

\subsection{Vector Search-Augmented AI (VS4AI)}
\label{sec:vs4ai}
We investigate how vector search leverage their powerful capabilities in semantic indexing and efficient retrieval to address the challenges of LLMs outlined in Section~\ref{sec:challenge_ai}.

Vector search systematically addresses core LLMs challenges through RAG, which strategically integrates dynamic, external knowledge sources into the generative process~\cite{fan2024survey,zhao2024retrieval}.
The RAG framework has evolved from Naive RAG to Advanced RAG, and finally to Modular RAG~\cite{gao2023retrieval}.

\subsubsection{\textbf{Naive RAG (RAG 1.0)}}
The core innovation of the Naive RAG phase was establishing a stable, fixed, linear framework for injecting external, non-parametric knowledge into the generation process~\cite{lewis2020retrieval,karpukhin2020dense,izacard2020distilling}.
The Naive RAG follows a foundational pipeline that includes indexing, retrieval, and generation~\cite{gao2023retrieval}.
\begin{itemize}
      \item \underline{\emph{Indexing}}: In this preparatory phase, a corpus of external knowledge (e.g., documents, web pages) is partitioned into chunks. These chunks are then encoded into vector representations using an embedding model, constructing an index that enables efficient similarity comparisons.
      \item \underline{\emph{Retrieval}}: Upon receiving a user query, it is similarly encoded into a vector. The system then performs a similarity search over the pre-built index to identify and retrieve the top-k most relevant text chunks.
      \item \underline{\emph{Generation}}: Finally, the retrieved contexts and the original query are concatenated into a prompt, which is fed into a LLM. The LLM is then tasked with synthesizing a fluent and coherent answer based on the provided context.
\end{itemize}

The Naive RAG pipeline successfully demonstrated that augmenting LLMs with retrieval significantly enhances factual accuracy and reliability~\cite{oche2025systematic}.

\subsubsection{\textbf{Advanced RAG (RAG 2.0)}}

The Naive RAG pipeline, despite its success, suffered from a structural vulnerability: it was highly susceptible to the retrieval quality.
The Advanced RAG framework focuses on optimizing the individual components of the pipeline, particularly the pre-retrieval, retrieval, and post-retrieval modules, to maximize the quality of the context input.
\begin{itemize}
      \item \underline{\emph{Pre-retrieval enhancement}}: 
      This line of work aims to refine the index structure or the query itself to yield better search results. 
      \emph{Index optimization} transforms the way knowledge is organized within RAG systems to enable more precise information retrieval. 
      Key approaches include adaptive chunking strategies that preserve semantic coherence~\cite{merola2025reconstructing}, strategic metadata attachments that enrich document representation~\cite{gollapudi2023filtered,cai2024navigating}, and knowledge graph enhancements that capture complex semantic relationships~\cite{wang2024knowledge,liang2025kag}.
      \emph{Query optimization} focuses on refining user queries to better express information needs and align with retrieval system characteristics.
      This encompasses query rewriting, where LLMs reformulate original queries into more effective versions~\cite{ma2023query,peng2024large,gao2023precise}, and query expansion, which generates multiple related queries to retrieve broader sets of relevant documents~\cite{zhou2022least,jagerman2023query}.
      \item \underline{\emph{Retrieval enhancement}}: \emph{Hybrid retrieval}~\cite{sawarkar2024blended}, which integrates sparse and dense methods, significantly enhances robustness of RAG for complex real-world queries by leveraging their complementary strengths.
      Sparse retrieval ~\cite{robertson2009probabilistic} ensures precise lexical matching but struggles with semantic variations, while dense retrieval~\cite{karpukhin2020dense} captures semantic similarity yet may miss critical keyword matches.
      \emph{Fusion function} determines the final unified ranking from the heterogeneous scores of the sparse and dense retrievers~\cite{bruch2023analysis,rackauckas2024rag}.
      \emph{Multi-Vector models} represent a crucial evolution within the hybrid retrieval paradigm~\cite{santhanam2022colbertv2,engels2023dessert}. 
      By representing documents and queries with multiple fine-grained vectors instead of a single dense embedding, these models enable more nuanced semantic matching~\cite{bian2025igp,jayaram2024muvera,nardini2024efficient}.
      \item \underline{\emph{Post-retrieval enhancement}}: 
      After the initial retrieval of candidate passages, a critical post-processing phase is applied to refine the context and integrate it effectively with the query before generation.
      \emph{Re-ranking} refines the retrieved results in RAG to ensure that only the most relevant documents are passed to the generator~\cite{dong2024don,yu2024rankrag}. 
      By reordering candidates based on semantic relevance and query alignment, it effectively reduces noise, improves factuality, and enhances the overall quality of generated responses.
      \emph{Context compression} mitigates the limited context window of large language models by condensing retrieved text into shorter, information-dense inputs~\cite{cheng2024xrag,rau2024context,fang2025attentionrag}. 
      This step improves efficiency, lowers cost, and allows RAG systems to scale to larger corpora and longer retrieval contexts.
\end{itemize}

\subsubsection{\textbf{Modular RAG (RAG 3.0)}}
Modular RAG represents the evolution of RAG beyond fixed, linear pipelines (Naive RAG) or component-level optimization (Advanced RAG), introducing architectural flexibility to the RAG framework.
This modular design enables dynamic workflow composition and specialized optimization, enhancing performance on complex tasks through flexible component integration and replacement.
\begin{itemize}
    \item \underline{\emph{Core Modules}}:
    Modular RAG develops specialized components that replace linear pipelines. 
    The \emph{Router} dynamically selects the processing paths~\cite{li2023classification}. 
    The \emph{Memory} module preserves contextual history for multi-turn interactions~\cite{cheng2023lift}. 
    The \emph{Predictor} generates hypothetical contexts to improve retrieval~\cite{yu2022generate}. 
    The \emph{Search} module enables targeted queries across diverse data sources~\cite{wang2023knowledgpt}.
    \item \underline{\emph{Advanced Workflow Patterns}}:
    Beyond new modules, Modular RAG explores customize composition of these modules into non-linear workflows.
    \emph{Iterative} patterns enable multi-step reasoning loops~\cite{shao2023enhancing}.
    \emph{Demonstrate-search-predict} patterns leverage example-driven retrieval with proactive context generation~\cite{khattab2022demonstrate}.
    \emph{Conditional} patterns allow systems to adaptively invoke modules based on query needs~\cite{asai2024self}.
\end{itemize}

\subsection{Closing the Loop: End-to-End Optimization}
While RAG has demonstrated the power of combining vector search with LLMs, their retriever and generator are optimized separately with divergent objectives. 
The retriever maximizes query-document similarity, while the generator focuses on text likelihood. 
This objective misalignment leads to suboptimal coordination.

We analyze recent research on end-to-end optimization, which co-trains the retriever and generator within a unified framework to fully unlock the potential of the ``virtuous cycle".

\begin{itemize}
    \item \underline{\emph{Differentiable Retrieval}}:
    Unlike traditional retrieval that uses a fixed, non-differentiable index, differentiable retrieval enables gradients from the generator's loss to flow back and optimize the retriever~\cite{anonymous2025drag,zhao2023differentiable}.
    \item \underline{\emph{Joint Retriever–Generator Training}}:
    Instead of making retrieval differentiable, some methods coordinate retriever and generator updates so both are optimized toward the same end-task (e.g., QA accuracy)~\cite{singh2021end}.
    The retriever may still select discrete top-k documents, and gradients may not flow through retrieval directly.
    Instead, the model is updated alternately or jointly through indirect signals.
\end{itemize}

\subsection{Challenges and Opportunities}

\textbf{Self-Evolving Indexes.}
Most learning-to-index methods operate under a static paradigm: trained once on a fixed dataset and deployed without further adaptation. 
However, real-world vector databases are inherently dynamic: embeddings drift as upstream models update, and data distributions continuously shift. Consequently, it becomes essential to develop continual index learners capable of incrementally adapting partition boundaries, hash functions, or quantization codes without requiring full retraining. 
Such adaptive systems would maintain retrieval performance under data streams while preserving backward compatibility with existing index structures.

\textbf{Retrieval Memoization.}
RAG systems face critical challenges, including high latency caused by repeated retrieval cycles, increased computational costs due to redundant vector searches, and inconsistent responses to semantically similar queries.
These limitations highlight a pressing need for an effective RAG caching mechanism, which mitigates these inefficiencies by storing and intelligently reusing previously retrieved or processed results.
By avoiding unnecessary retrieval and generation steps, RAG caching substantially reduces response time and computational overhead, while enhancing output stability and reproducibility.

\textbf{Autonomous execution.}
Modular RAG framework faces fundamental limitations due to its reliance on predefined workflows and lack of dynamic reasoning. 
They cannot replan on the basis of intermediate results or self-evaluate information sufficiency. 
This requires the evolution to Agentic RAG, which introduces dynamic planning, continuous self-monitoring, and proactive tool usage. 
By functioning as an autonomous agent that can formulate and execute multi-step strategies, Agentic RAG transforms the system from a stateless executor into a cognitive assistant capable of handling complex, ambiguous tasks through strategic reasoning and adaptive learning.

\section{Presenters}

\textbf{Jiuqi Wei} is a researcher of Oceanbase Lab, Ant Group.
He received his Ph.D. degree at Institute of Computing Technology, Chinese Academy of Sciences.
His research interests include vector database, hybrid retrieval, and data management.
His recent works~\cite{detlsh,wei2025subspace,wei2025dominate,wei2023data,li2024disauth,zhuang2024cbcms,fu2024securing,fu2023ti,fei2023flexauth,zhang2025polaris,zhuang2026structured,song2026csattention} concentrate on approximate nearest neighbor search and data governance.

\textbf{Quanqing Xu} is the technical director of Oceanbase Lab, Ant Group.
His research interests primarily include distributed database systems and storage systems~\cite{yang2025lcl+,fang2025malt,hu2025oltp,xu2025oceanbase,luo2025answer,luo2025rm2,yang2022oceanbase,yang2023oceanbase,zhang2023efficient,qu2022current,xu2024native,han2024palf,yang2023lcl}.
He is a Fellow of the IET, a distinguished member of CCF, a senior member of ACM and IEEE.

\textbf{Chuanhui Yang} is the CTO of OceanBase, Ant Group.
His research focuses on database, distributed
systems, and AI infra~\cite{yang2022oceanbase,yang2023oceanbase,zhang2023efficient,qu2022current,xu2024native,han2024palf,fang2025malt,hu2025oltp,xu2025oceanbase,luo2025answer,luo2025rm2}.
As one of the founding members, he led the previous architecture design, and technology research and development of OceanBase, realizing the full implementation of OceanBase in Ant
Group from scratch.

\section*{AI-Generated Content Acknowledgement}
In the research process for this work, the authors used large language models (Gemini and ChatGPT) for text polishing and writing assistance. 
After using these tools, the authors reviewed and edited the content as needed and take full responsibility for the publication.

\bibliographystyle{IEEEtran}

\bibliography{IEEEabrv,ref}

\end{document}